\documentclass[runningheads]{llncs}

\usepackage[T1]{fontenc}
\usepackage{graphicx}
\usepackage{xurl}
\usepackage{amsfonts}

\usepackage{booktabs}
\usepackage{tabularx}
\usepackage[table]{xcolor}
\usepackage{threeparttable}
\begin{document}
\raggedbottom

\title{Quantifying Observable High-Frequency Swapping on Arbitrum}





\author{Shijian Chen\inst{1}
\and Ya Chen\inst{2}
\and Jing Cai\inst{3}
\and Catherine Liu\inst{4}\thanks{Corresponding author.}}

\authorrunning{S. Chen et al.}

\institute{
The Hong Kong Polytechnic University, Hong Kong, China\\
\email{24056659r\@connect.polyu.edu.hk}
\and
Hefei University of Technology, Hefei, China\\
\email{ychen\@hfut.edu.cn}
\and
The Hong Kong Polytechnic University, Hong Kong, China\\
\email{jing.cai\@polyu.edu.hk}
\and
The Hong Kong Polytechnic University, Hong Kong, China\\
\email{macliu\@polyu.edu.hk}
}



\maketitle             
\begin{abstract}
Layer-2 rollups have reshaped Ethereum’s transaction economy by replacing mempool competition with deterministic sequencing, sub-second block times, and negligible gas fees. While these properties suppress classical Miner/Maximal Extractable Value (MEV), they give rise to a new and previously unrecognized behavioral regime. In this paper, we introduce the concept of \textit{High-Frequency Swapping (HFS)}, referring to continuous single-hop swaps at machine cadence within decentralized exchanges. Despite its growing footprint, this phenomenon has not been systematically identified or quantified in prior work. We conduct a large-scale measurement of HFS on Arbitrum, using 18 months of full-chain data (Jan.~2023--Jun.~2024). Specifically, we construct a reproducible pipeline that isolates voluntary single-swap transactions, attributes them to the swapper level, and classifies HFS behavior through inter-arrival dynamics. The analysis uncovers 477 distinct HFS swappers responsible for nearly 30 million swaps and over \$10$^{11}$ in notional volume. Our study conducts a comprehensive empirical investigation of HFS from multiple perspectives. We begin with a global overview of activity patterns, and then examine temporal dynamics, swapper identity, token coverage, and venue concentration, and explore swap size, time gap, and order direction. Finally, we explore case-level behavior, including stablecoin arbitrage, CEX–DEX execution gaps, and short-horizon round-trip trading. Across these dimensions, we identify consistent structural regularities that distinguish HFS from conventional retail or arbitrage activity. This work provides a large-scale address-level empirical characterization of \textit{High-Frequency Swapping} on Arbitrum and a broad empirical foundation for future research on decentralized market microstructure.

\end{abstract}

\section{Introduction}
\label{sec: introduction}

Decentralized exchanges (DEXs) on modern blockchains host a rapidly growing share of global crypto trading activity~\cite{xu2023sok,malamud2017decentralized}. Unlike centralized exchanges, their transparency and composability expose every swap, liquidity adjustment, and arbitrage loop directly on-chain~\cite{hagele2024centralized}. This visibility has enabled a decade of research on market microstructure and Miner/Maximal Extractable Value (MEV)~\cite{daian2019flash,qin2022quantifying,qin2021empirical,zhou2021high,li2023demystifying,yang2024sok,wang2022cyclic,mclaughlin2023large,moallemi2024analysis,qin2023mitigating,heimbach2022eliminating,zhou2021high,zhou2023sok}. On Ethereum mainnet (L1), MEV studies have shown how miners or builders exploit transaction ordering and mempool access to capture intra-block profits. These behaviors, such as sandwich attacks~\cite{heimbach2022eliminating,zhou2021high,zhou2023sok}, cyclic arbitrage~\cite{wang2022cyclic,mclaughlin2023large}, and liquidation sniping~\cite{qin2021empirical,moallemi2024analysis,qin2023mitigating}, dominate the literature and shape our current understanding of on-chain trading.

However, Ethereum’s Layer~2 (L2) rollups have quietly altered this logic. Arbitrum~\cite{kalodner2018arbitrum}, the largest L2 network by total value secured, executes transactions under a single sequencer with deterministic ordering, sub-second block intervals, and negligible gas fees. This architecture removes classical mempool-based MEV opportunities but simultaneously enables a new behavioral regime: addresses that interact with DEXs at machine-level cadence. These activities, which we term \textit{High-Frequency Swapping (HFS)}, represent a distinct form of on-chain execution where the strategic edge lies not in manipulating transaction order, but in reacting faster and more continuously within deterministic markets.

Despite abundant research on L1 MEV and cross-venue arbitrage, there is no systematic characterization of sustained continuous swapping on L2 rollups. Existing studies either aggregate activity across tokens and venues or focus on atomic arbitrage bundles, overlooking address-level microdynamics that emerge when latency, rather than ordering, becomes the scarce resource. We therefore ask: who are the actors behind these rapid swaps, how persistent are their temporal patterns, and do they obtain consistent execution advantages and short-horizon profitability?

This paper provides the first systematic characterization of HFS on Arbitrum. Using 18 months of full-chain data (Jan.~2023--Jun.~2024), we build a reproducible pipeline that isolates voluntary single-swap transactions, attributes swaps at the address level, and reconstructs inter-arrival dynamics. We identify 477 distinct HFS swappers executing nearly 30 million swaps with over \$10$^{11}$ in notional volume. We then analyze temporal cadence, identity distribution, token and venue coverage, and swap directionality, and present case studies showing stablecoin-pair specialization, CEX--DEX alignment, and recurrent round-trip patterns. Our contributions are as follows,
\begin{itemize}
    \item \textbf{First characterization of HFS} To the best of our knowledge, we are the first to introduce and empirically validate High-Frequency Swapping as a distinct category of on-chain behavior, showing that it accounts for a substantial share of Arbitrum’s DEX volume.
    \item \textbf{Methodology for systematic detection.} We present a scalable pipeline for identifying sustained continuous swapping from full-chain traces and release datasets to support follow-up research.
    \item \textbf{Statistical and behavioral exploration.} We provide a multi-level empirical analysis of HFS, covering aggregate scale and temporal dynamics, entity-level identities and venue concentration, and microstructural patterns such as inter-swap gaps, trade sizes, and directional alternation. We further analyze representative cases, stablecoin arbitrage, CEX--DEX execution alignment, and round-trip trading.
\end{itemize}

\section{Background and Related Work}
\label{sec: background}




\subsection{Blockchain, Ethereum, Arbitrum} Blockchain provides a decentralized and transparent infrastructure for recording transactions, supporting applications such as cryptocurrencies, decentralized finance, and digital asset management~\cite{chen2024exploring}. Ethereum has become the dominant smart-contract platform, yet it still faces scalability bottlenecks, including limited throughput and high transaction fees. The ecosystem has increasingly adopted Layer-2 solutions, with rollups being the most widely used. Arbitrum~\cite{kalodner2018arbitrum}, one of the leading rollups, batches and compresses L2 transactions before posting proofs to L1, significantly reducing gas costs and improving scalability. In practice, average swap fees on Arbitrum are typically below \$0.05, and block times are around 250 ms, compared with Ethereum’s 12-second block interval. As of Aug.~27,~2025, L2Beat reports that Arbitrum is the largest L2 network, with Total Value Secured of \$19.4B~\cite{l2beat_arb}.

\subsection{Miner/Maximal Extractable Value}
A large body of work has investigated Miner/Maximal Extractable Value (MEV) in Ethereum~\cite{daian2019flash,qin2022quantifying,qin2021empirical,zhou2021high,li2023demystifying,yang2024sok,zhou2023sok,mclaughlin2023large}. The concept of MEV originated from early studies on Ethereum's proof-of-work (PoW) era~\cite{buterin2016ethereum}, where miners could arbitrarily reorder, insert, or censor transactions within the blocks they produced. Daian et al.'s seminal work~\cite{daian2019flash} first revealed that decentralized exchanges on Ethereum are inherently vulnerable to frontrunning and backrunning attacks, as miners can exploit mempool transparency to capture arbitrage opportunities before ordinary users. Subsequent empirical analyses demonstrated that such behavior was not exceptional but systematic, giving rise to persistent categories of extractable value, including sandwich attacks~\cite{heimbach2022eliminating,zhou2021high,zhou2023sok}, cyclic arbitrage~\cite{wang2022cyclic,mclaughlin2023large}, and liquidation sniping~\cite{qin2021empirical,moallemi2024analysis,qin2023mitigating}.

Following Ethereum’s transition to proof-of-stake (PoS) in the Merge~\cite{liu2025ethereum}, block production shifted from miners to validators, but incentives for transaction reordering remained. To separate consensus from execution and reduce validator-level manipulation, Ethereum introduced the Proposer--Builder Separation (PBS) model~\cite{heimbach2023ethereum}. Under PBS, specialized builders assemble transaction bundles and bid for block inclusion via relays, while validators (proposers) select the highest-paying block~\cite{wahrstatter2023time}. This design improved efficiency and reduced in-protocol censorship, but it also reshaped MEV extraction into a professionalized market dominated by a few industrial-scale builders~\cite{oz2024wins,yang2025decentralization}. Recent measurements~\cite{mevboost} show that more than 80\% of Ethereum blocks are built through MEV-Boost~\cite{mevboost1}, an open-source relay system that operationalizes PBS. Together, these developments show how Ethereum’s transaction economy evolved from localized mempool competition into an institutionalized layer of value extraction within the block-construction supply chain.

Parallel to empirical MEV studies, AMM microstructure theory formalizes Loss-versus-Rebalancing (LVR) as the adverse-selection cost incurred when AMM prices lag external reference markets~\cite{milionis2022automated,milionis2024automated}. This line of work establishes CEX--DEX price-discrepancy arbitrage as a core mechanism in AMM-based trading. Our study is complementary: rather than re-deriving LVR theory, we provide a large-scale address-level empirical characterization of sustained high-frequency swapping behavior on Arbitrum over an 18-month horizon.

Despite extensive understanding of MEV on Ethereum, most existing analyses remain focused on L1 dynamics, where transaction ordering and inclusion privileges are the main sources of extractable value. In contrast, L2 rollups such as Arbitrum operate under a different execution paradigm: transactions are sequenced deterministically by a single sequencer, without public mempool competition or multi-party bidding. As a result, classical frontrunning, backrunning, and sandwich attacks become largely infeasible~\cite{ferreira2024rolling}. Even so, incentives for rapid response and micro-level arbitrage persist, now appearing as continuous high-frequency interactions within automated market maker (AMM) pools. This shift reframes MEV from an ordering-based to a timing-based phenomenon, where the edge lies in reacting faster to price movements within a deterministic L2 environment.

\section{Motivation Example: 0x2d512b}
\label{sec: motivation_example}

We begin with a concrete case study that illustrates the scale and style of HFS on Arbitrum. 0x2d512b~\cite{addr_0x2d512B} emerges as an extreme outlier within a single Uniswap V3 ARB/USDC pool (0xb0f6ca~\cite{pool_0xb0f6ca}), it executed 127,296 swaps between block 100,321,269 and 123,441,868. During the same period, the pool recorded a total of 258,897 swaps, meaning that a single address was responsible for nearly half (49.2\%) of all activity. Such concentration is far beyond retail behavior and provides direct evidence of institutional-scale HFS activity. The swapping burst was seeded by an initial transfer of 200{,}000 `bridged USDC.e'~\cite{token_usdce} from 0xabcd71~\cite{addr_0xabcd71}, which converted into 199,797.41 USDC~\cite{token_usdc} as the starting balance (with no ARB holdings). Every transaction contained exactly one atomic swap, without routing or multi-hop interactions. At the end of the burst, the account’s final holdings --- 131,946 ARB \& 135,192 USDC --- were returned in the same quantities to 0xabcd71~\cite{addr_0xabcd71}, strongly suggesting that 0x2d512b~\cite{addr_0x2d512B} functioned as an operational sub-account or delegated trading wallet.  Apart from its uninterrupted swapping, 0x2d512b~\cite{addr_0x2d512B} was involved in only five direct USDC transfers~\cite{txn_0xb1f880,txn_0x9517a1,txn_0x1e9728,txn_0x7e56f3,txn_0x91601f}, receiving 50,000 USDC from another HFS swapper 0x769d6a\cite{addr_0x769D6a} and 16,662 USDC from 0xabcd71\cite{addr_0xabcd71}, and sent three transfers (13,333 USDC, 25,000 USDC, and 22,000 USDC) to 0xE27AE1\cite{addr_0xE27AE1}. These scattered transfers highlight the account’s otherwise singular behavior, i.e., nearly all of its on-chain actions were repetitive swaps within a single pool, with only minimal off-pool fund movements. 

\begin{figure}[htbp]
\centering
\includegraphics[width=1\linewidth]{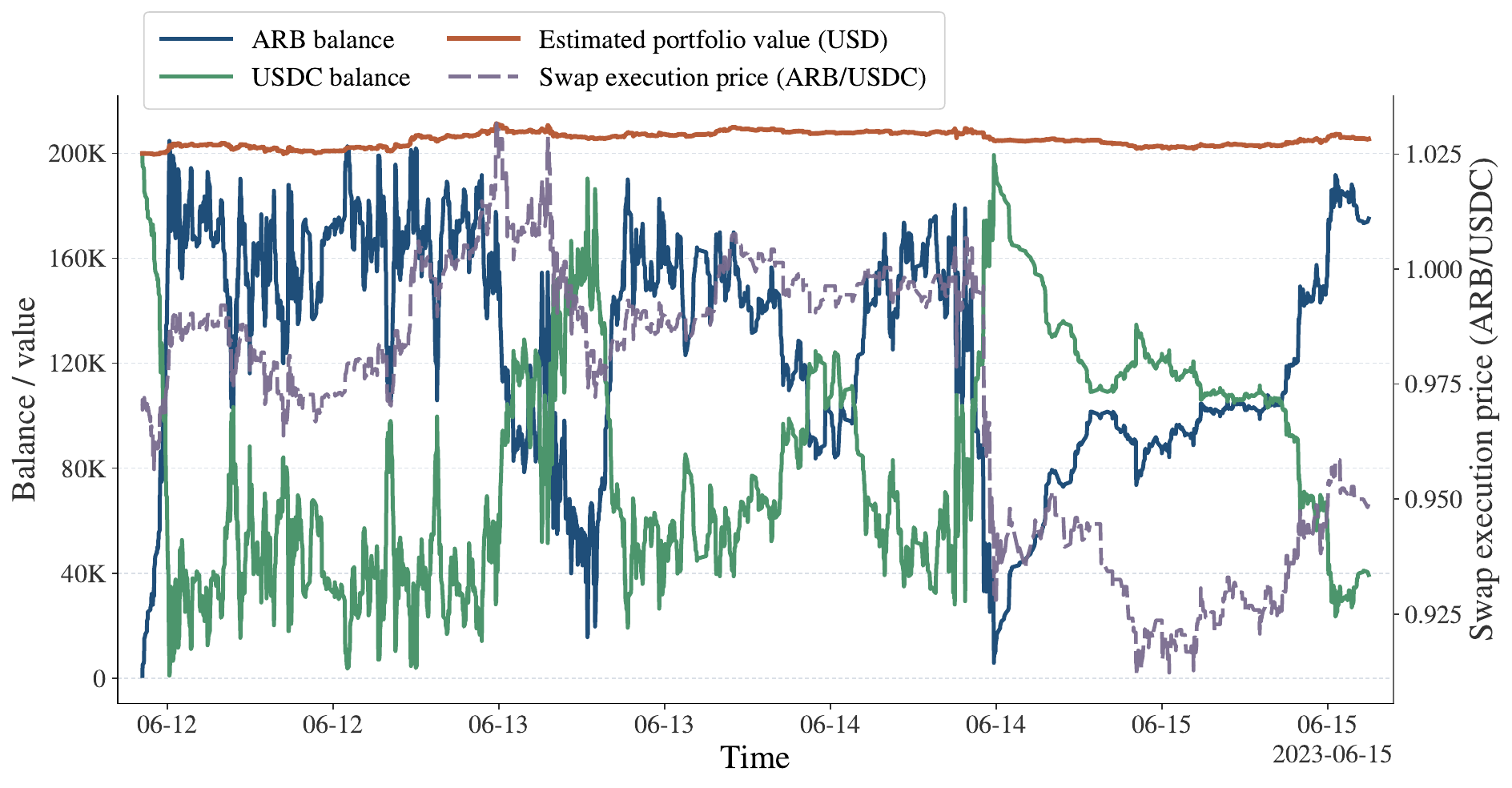}
\caption{ARB \& USDC balances, estimated USD value, and swap execution price of swapper 0x2d512b in the Uniswap V3 ARB/USDC pool. Execution price is defined as the effective USDC/ARB rate of each swap.}
\label{fig:example}
\end{figure}

0x2d512b~\cite{addr_0x2d512B} spent 5.846 ETH across all swaps, meaning that sustaining such intensity carried tangible execution costs. At the same time, if we mark the final portfolio to the ARB price observed in the last swap and combine it with the USDC holdings, the address shows an overall profit of around \$20,000. Figure~\ref{fig:example} plots its ARB and USDC balances, the estimated portfolio value in USD, and the execution price trajectory from Jun.~12,~2023, 07:11:02~UTC to Jun.~16,~2023, 00:00~UTC\footnote{We restrict the plot to this interval for clarity, as plotting the entire activity window would obscure patterns with excessive density.}. The pattern is one of relentless, one-sided bursts of swaps rather than opportunistic alternation of buys and sells. This case study highlights two insights. First, it demonstrates that HFS swappers on Arbitrum can be empirically identified through address-level traces. Second, it illustrates that their execution style diverges sharply from Ethereum L1 searchers, e.g., adversarial MEV tactics in the absence of a public mempool and PBS auctions. This motivational example motivates the broader measurement study that follows.

\section{Methodology}
\label{sec: methodology}

In this section, we design a pipeline to systematically identify and analyze 18-month scale HFS swappers on Arbitrum. The methodology proceeds in three logical stages: transaction filtering, swapper attribution, and high-frequency classification, followed by a note on data collection.

\textbf{1) Transaction Filtering.} We begin from the full set of decentralized exchange (DEX) transactions on Arbitrum and apply structural filters to isolate economically meaningful single-swap events. 
\begin{itemize}
    \item \textit{Single-swap rule. }We retain only transactions containing exactly one swap event, thereby excluding multi-hop routing, cyclic arbitrage, and other complex interactions that obscure the intent of a single swapper.
    \item \textit{Exclusion of involuntary trades. }Transactions triggered by liquidation mechanisms are removed, since they reflect forced position closures rather than voluntary trading behavior.
    \item \textit{CEX-token focus. } To emphasize trades with clear cross-venue arbitrage potential, we restrict attention to tokens that were listed on Binance spot or perpetual futures markets on or before Jul.~2024 and that have active ERC-20 contract addresses deployed on Arbitrum, ensuring that all analyzed pairs are simultaneously (i) liquid on centralized exchanges and (ii) tradable on Arbitrum.
\end{itemize}
Taken together, these filters yield a high-confidence subset of economically relevant single-swap transactions. Importantly, they also exclude the main categories of MEV behavior documented on Ethereum L1. Liquidation-driven swaps are removed, cyclic arbitrage typically requires at least two sequential swaps~\cite{mclaughlin2023large}, and sandwich attacks, while prevalent on Ethereum L1, are widely believed to be infeasible~\cite{ferreira2024rolling} on Arbitrum given its sequencer-based First-Come First-Served ordering~\cite{kalodner2018arbitrum}. Thus, our design provides an operationally clean lens on voluntary swapper activity without conflating it with adversarial MEV strategies.

\textbf{2) Swapper Attribution.} We define the swapper as the address recorded as the recipient of output tokens in swap event logs (e.g., the recipient field in Uniswap V3’s Swap event). To avoid misattribution, we exclude infrastructure contracts such as routers and aggregators, which may appear as recipients but do not represent the actual economic actors. This approach ensures attribution to genuine accounts that acquire tokens through swaps.

\textbf{3) High-Frequency Classification.} To distinguish HFS swappers from casual users, we analyze the temporal rhythm of transactions for each address.  For every swapper $u$, we order its swaps by block timestamp $\{t_1, t_2, \dots, t_N\}$ and compute inter-arrival gaps as $\mathrm{gap}_i = t_i - t_{i-1}$ for $i=2,\dots,N$. From these gaps, we extract summary statistics including the median gap $\tilde{g}$, the 90th percentile gap $g_{0.9}$, the maximum observed gap, and the fraction of long gaps (e.g., exceeding one hour). We classify $u$ as a \emph{HFS swapper} if it satisfies all of the following conditions: $N \geq 150$, $\tilde{g} \leq 120$ seconds, $g_{0.9} \leq 900$ seconds, and $p_{>1h} \leq 0.20$. These thresholds are calibrated manually from the empirical distribution of inter-arrival times, striking a balance between capturing sustained automated activity and avoiding false positives from occasional bursts. Importantly, we do not require bi-directionality (both buys and sells of the same token), since many high-frequency strategies on L2 are observed to be strongly one-sided. This rule is conservative and excludes most retail-like users who trade intermittently.

\textbf{4) Data Collection. }We construct a transaction-level dataset of swaps on Arbitrum between Jan.~2023 and Jun.~2024. Our pipeline begins from raw execution traces of the Arbitrum chain, where we parse \textit{Swap()} events emitted by major decentralized exchanges, including Uniswap v2/v3, SushiSwap, Camelot, PancakeSwap, Ramses, Trader Joe, and etc. For each transaction, we extract the traded token pair, input and output amounts, and attribute the swap to the account recorded as the recipient in the \textit{Swap()} event.

\section{Empirical Results}
\label{sec: result}

\subsection{Overview} 
Our high-frequency filter identifies $29.9$M single-swap transactions on Arbitrum between Jan.~2023 and Jun.~2024, executed by $477$ distinct swappers. This subset accounts for about $22\%$ of all observed swaps in the same period, underscoring that high-frequency activity is a substantial share of overall market flow. The aggregate notional exceeds $1.02\times 10^{11}$ USD, highlighting systemic scale. While most trades are small (median \$1.75k), the distribution is heavy-tailed (p90 \$7.9k; p99 \$22.4k).

\subsection{Temporal Dynamics} 

\begin{figure}[t]
    \centering
    \includegraphics[width=1\linewidth]{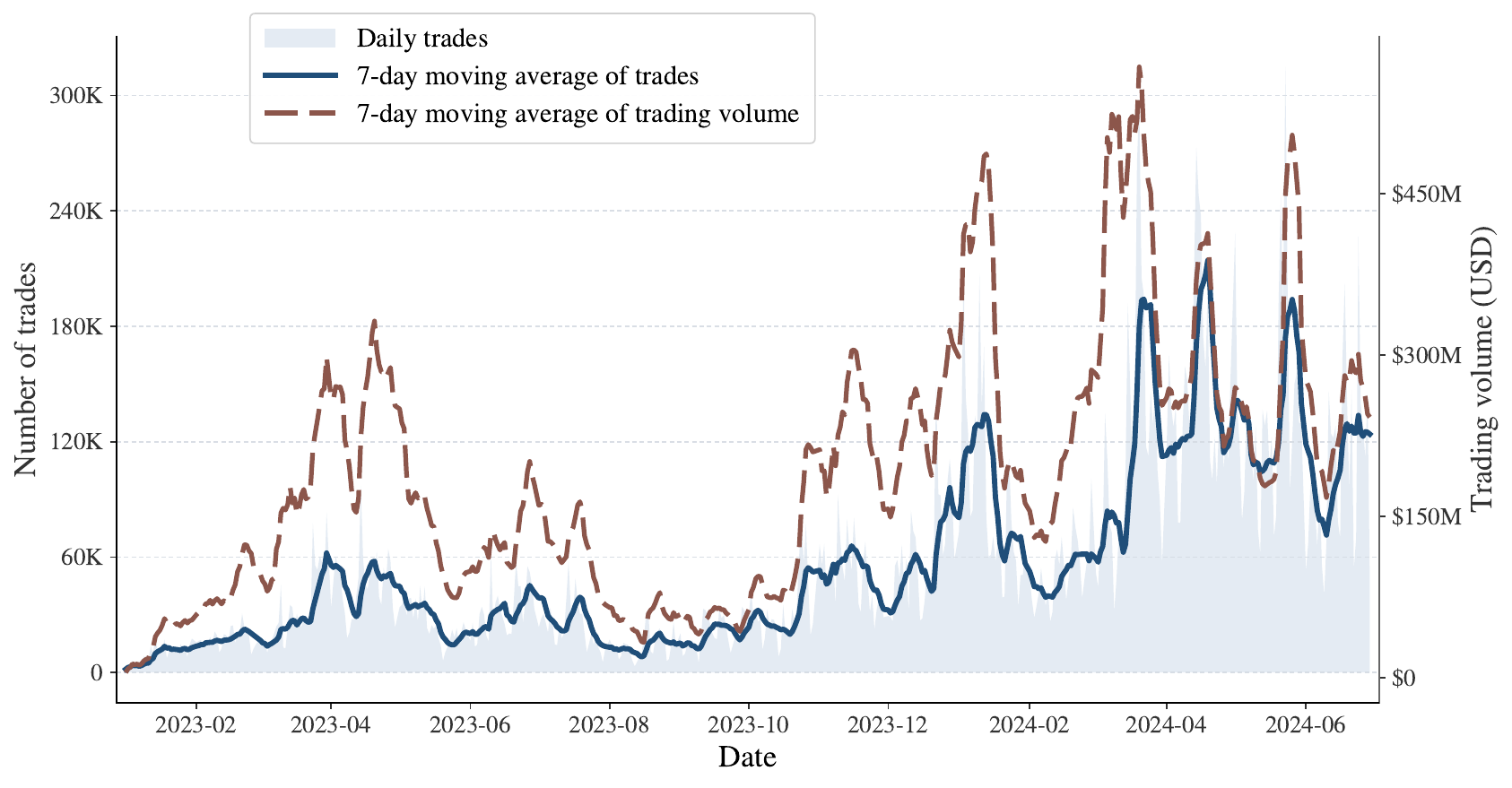}
    \caption{Monthly evolution of high-frequency DEX trades and notional volume in USD from Jan.~2023 to Jun.~2024, revealing strong cyclical fluctuations and a sustained growth trajectory over 18 months.}
    \label{fig1}
\end{figure}

Among all days in our 18-month sample, May~23,~2024, stands out as the single-day peak with over 314,842 transactions and more than \$1.08B notional volume, concentrated across only 57 active swappers. Figure~\ref{fig1} summarizes the monthly evolution of HFS swapper activity between Jan.~2023 and Jun.~2024. The market footprint expanded rapidly in the first half of 2023: trades increased from 291,742 in January to over 1,322,581 in April, accompanied by notional volumes rising from \$1.17B to over \$7.20B. Activity softened in mid-2023, bottoming in August with only 424,923 trades and \$1.70B notional, before rebounding into Q4, where November–December recorded 1,504,064 to 1,860,770 trades and more than \$6.67B to \$7.56B volume. The most pronounced acceleration occurred in early Jan.~2024, accounting for 2,760,604 trades and over \$9.10B volume, nearly tripling mid-2023 levels. Mar.~2024 recorded 3,483,929 trades and \$12.86B notional, while Apr.~2024 marked the sample-period peak with 4,242,892 trades. Overall, the 18-month panel reveals a cyclical pattern: alternating contractions and expansions, yet with a persistent upward drift in both trades and notional turnover. 

A striking feature is the dominance of ETH-related swapping pairs: USDC/WETH and WETH/USDC alone contributed over half of daily volume (52.29\%), while USDT0/WETH and WETH/USDT0 jointly added another 19.56\%. This concentration indicates that high-frequency strategies were overwhelmingly focused on ETH price action rather than broad-based arbitrage. The timing aligns with major market-moving news. On May~20,~2024, Bloomberg ETF analyst Eric Balchunas reported that the SEC was increasingly likely to approve spot Ethereum ETFs, raising estimated approval odds from 25\% to 75\%. This rumor sparked an immediate price surge in ETH from roughly \$3,150 to above \$3,800 within a single day, and momentum carried into the official SEC decision deadline of May~23. During this window, ETH markets experienced unusually high volatility and liquidity demand, providing fertile ground for HFS arbitrage and liquidity-provision strategies. The combination of exogenous regulatory news and endogenous liquidity dynamics thus explains why ETH-related swapping pairs dominate the observed peak. These patterns suggest that HFS swappers on Arbitrum respond not only to on-chain opportunities but also to off-chain macro events, with news shocks transmitting almost instantly into DEX microstructure.

\subsection{Swapper Identity}


The number of distinct HFS swappers on Arbitrum averages around 42 unique addresses per day, ranging between 17 and 70. This consistent daily activity indicates that HFS is primarily driven by a persistent core group of addresses that operate nearly every day, likely professional or institutional actors, rather than a constantly rotating set of opportunistic entrants. We cross-referenced all 477 HFS swapper addresses against Arkham Intelligence labels~\cite{arkm}. Only 8 addresses (1.68\%) could be linked to professional trading firms, including Wintermute (2)~\cite{wintermute}, Flow Traders (1)~\cite{flowtraders}, Selini Capital (1)~\cite{selini}, and Manifold Trading (4)~\cite{manifold}. From this perspective, institutional actors are the only ones that can be reliably identified through labels. For instance, Manifold Trading~\cite{manifold} is a quantitative investment firm reported to focus on systematic strategies in the crypto ecosystem. Another 27 addresses (5.66\%) carried weak identifiers, such as ENS names (e.g., \texttt{drgroove.eth}, \texttt{lemon.eth}, \texttt{daoid.eth}) or OpenSea usernames (e.g., ``Died-Fanboy-123''), which provide limited attribution value.  The overwhelming majority, 442 addresses (92.66\%), remained unlabeled or associated with non-informative tags, making it impossible to infer ownership or coordination. This distribution suggests that if coordinated behavior exists among HFS swappers, it is not explicitly documented through public labels.

\subsection{Token \& Venues Coverage}
\subsubsection{Token Coverage}
\begin{figure}[t]
    \centering
    \includegraphics[width=0.95\linewidth]{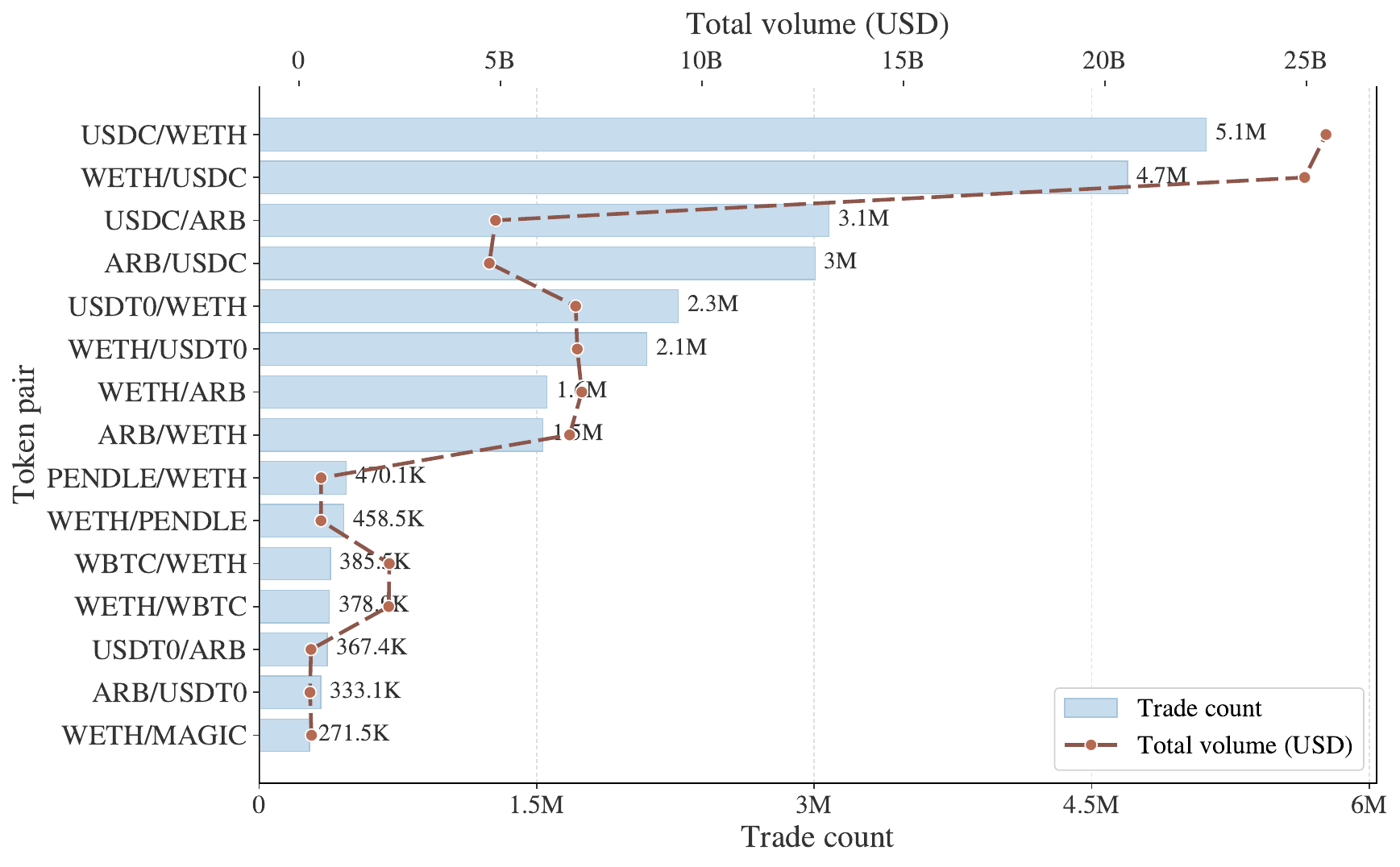}

    \caption{Top traded token pairs: the comparison highlights the near-symmetry of buy/sell directions and the dominance of ETH-based pairs across both dimensions.}
    \label{fig:top_pairs}
\end{figure}

High-frequency activity on Arbitrum is overwhelmingly concentrated in a handful of token pairs. The most prominent are stablecoin–ETH pairs, with USDC-WETH alone contributing over 9,813,743 trades and a combined notional exceeding \$50.44B. The ARB token, as Arbitrum’s native asset, also features prominently: ARB-USDC and WETH-ARB jointly account for more than 6,086,221 trades and roughly \$23.35B in turnover. Beyond these, secondary ecosystem tokens such as Pendle, Magic, and GMX also exhibit substantially high-frequency volumes in their WETH pairs, highlighting the extent to which specialized protocols attract algorithmic liquidity. Finally, the presence of WBTC pairs underscores that Bitcoin-backed liquidity is consistently integrated into this high-frequency segment despite Arbitrum being an Ethereum Layer-2 environment. As shown in Figure~\ref{fig:top_pairs}, for the most liquid pairs, such as WETH-USDC, we observe that both trade counts and notional volumes appear in nearly symmetric form across directions (e.g., USDC$\rightarrow$WETH vs.\ WETH$\rightarrow$USDC). This is consistent with the design of automated market makers, where every swap in one direction mechanically induces the potential for swaps in the opposite direction. High-frequency traders amplify this symmetry by rapidly switching between buy and sell directions within the same pool, yielding closely matched counts on both sides. Importantly, the numbers are not perfectly identical, suggesting that temporary imbalances, for example, bursts of buy pressure following market news, do tilt activity toward one direction but only transiently. This pattern indicates that HFS swappers are systematically engaged in both sides of the market, providing liquidity-like activity even while acting as swappers.

\subsubsection{Venues Coverage}
We find that activity is highly concentrated in a handful of venues when mapping high-frequency trades to their hosting projects. Uniswap alone accounts for over 21.98M trades and more than \$85.06B in notional volume, dwarfing other projects. The second tier includes Camelot, PancakeSwap, SushiSwap, and Ramses, each hosting between one and three M trades and in the range of \$2.5B to 6.5B volume. Beyond these, the long tail of projects contributes only marginally: Trader Joe and KyberSwap appear as niche hubs, while smaller venues such as Arbswap, Zyberswap, and Chronos collectively account for less than 0.2\% of overall activity. This sharp concentration highlights that institutional-scale HFS strategies are primarily deployed on top-tier liquidity venues with deep and continuous markets.

\subsection{Swapping Gap, Size \& Direction}
\subsubsection{Inter-arrival Gap}
We analyze the distribution of inter-arrival times between consecutive swaps across all HFS swappers. For each swapper~$s$ with $N_s$ trades, we define the individual inter-arrival gaps as $g_i^{(s)} = t_i^{(s)} - t_{i-1}^{(s)}$ for $i = 2,\dots,N_s$, where $t_i^{(s)}$ is the timestamp of the $i$-th trade. The typical interval between two swaps is extremely short: the median gap is 4~seconds, while $p_{90}=101$~seconds and $p_{99}=881$~seconds (about 15~minutes). These values indicate that most swaps occur in rapid succession, forming a near-continuous execution stream. 
The longest observed interval spans roughly 24.7 days, reflecting occasional inactivity but contributing little to the overall distribution. Overall, the results suggest that HFS swappers on Arbitrum sustain machine-level cadence with minimal downtime, consistent with automated strategies tuned for throughput rather than discretionary human activity. At the swapper level, we compute the median gap~$\tilde{g}^{(s)}$ and 90th-percentile gap~$g_{0.9}^{(s)}$ for each address, then summarize these across the population. The median of~$\tilde{g}^{(s)}$ is 4~seconds, and that of~$g_{0.9}^{(s)}$ is 96~seconds, identifying a core group of `always-on' entities with ultra-tight execution cycles. At the population median swapper, we find $\tilde{g}^{(s)}=22$~seconds and $g_{0.9}^{(s)}=369$~seconds, and even by the 75th percentile, most still operate within sub-hour rhythms. This heterogeneity indicates that while a small set of swappers dominate the fastest regime, the majority maintain regular, programmatic activity rather than sporadic, retail-style trading.

To quantify the irregularity of swapper activity beyond raw inter-arrival gaps, we compute the \emph{Fano factor}~\cite{fano1947ionization} of trade counts in fixed-length time windows. For each swapper~$s$, let $N_{s,w}$ denote the number of swaps in window~$w$, and define $F_s = \frac{\mathrm{Var}_w(N_{s,w})}{\mathbb{E}_w[N_{s,w}]}$. This statistic measures dispersion relative to a Poisson benchmark: $F=1$ corresponds to Poisson-like arrivals, while $F>1$ indicates over-dispersion (bursty clustering) under the same windowing scheme. Across swappers, the median Fano factor is 150.93 and the 90th percentile reaches 885.50. These values are orders of magnitude above the Poisson benchmark, indicating that swapper activity is highly clustered into bursts rather than uniformly spread in time. Because Fano magnitudes depend on window length, we interpret these levels as evidence of strong temporal burstiness rather than as a structural arrival model. Such burstiness is consistent with event-driven strategies (e.g., reacting to CEX listing announcements or local price dislocations), where swapping intensity sharply spikes in response to external signals. This pattern reinforces the interpretation of these addresses as algorithmic actors coordinating on latency-sensitive opportunities, as opposed to passive or retail participants.


\subsubsection{Trade Size}  
We next examine that the median trade size among high-frequency swaps is approximately \$1.75k, suggesting that the bulk of such activity is conducted through relatively small orders. However, the distribution exhibits a pronounced right skew: the 90th percentile reaches about \$7.95k, the 99th percentile rises to nearly \$22.5k, and the maximum single transaction exceeds \$1.0M. These results indicate that while high-frequency strategies are primarily expressed through a large number of modest trades, the aggregate footprint is amplified by a long tail of occasional large-scale executions, consistent with institutional-style activity layered on top of smaller, more frequent flows. We next aggregate activity at the swapper level across the 477 identified HFS swappers. The median swapper executes approximately 2,223 swaps over the observation window, while the 90th percentile swapper reaches nearly 134,222 swaps. In terms of cumulative notional volume, the median swapper accounts for only \$2.91M, whereas the 90th percentile exceeds \$337.35M. This heavy-tailed distribution indicates a sharp concentration of activity: a handful of addresses sustain extremely high trade counts and volumes, consistent with institutional-grade market-making desks, while the majority of swappers engage at modest but persistent frequencies. Such stratification mirrors the division between retail-style HFS and professional liquidity provision observed in traditional equity markets.

\subsubsection{Directional Alternation}
We next analyze whether HFS swappers alternate rapidly between buy and sell directions or instead sustain longer unidirectional runs. The median alternation rate is only 0.25, meaning that a typical swapper executes about four consecutive trades in the same direction before switching sides. Even at the 90th percentile, the alternation rate remains below 0.45, while the bottom decile shows purely one-sided behavior with no directional switches at all. The distribution of run lengths further supports this observation. The median run length across pairs is three trades, and the 90th percentile reaches eight, implying that most swappers perform short bursts of same-side execution. These findings suggest that HFS swappers on Arbitrum engage in clustered, batch-style execution windows---`buy, buy, buy' or `sell, sell, sell'---rather than alternating direction on a per-trade basis. 0x51c728~\cite{addr_0x51C728} exemplifies extreme directional persistence across its three main pairs, i.e., ARB/WETH, ARB/USDC, and ARB/USDT, where it executed over 700,000 swaps totaling \$1.23B in notional volume, yet never switched trade direction. Each run spans the full swapping sequence, implying continuous buying or selling over the entire period. Such one-sided behavior likely reflects either automated inventory rebalancing or a fixed-role arbitrage leg in coordinated strategies. 0x44c6c4~\cite{addr_0x44c6c4} shows a markedly different pattern, characterized by frequent but structured alternation between buy and sell sides. On the USDC/WETH pair alone, it executed over 1.5M swaps totaling \$4.85B in volume, with an alternation rate of 0.065. Median run length is six consecutive trades, extending to forty at the 90th percentile, while the longest single-direction run spans 485 trades. This mix of repeated alternation and extended bursts might suggest an active liquidity-provision strategy, and continuously adjusting inventory while maintaining directional phases aligned with short-term market trends.

\section{Discussion}
\label{sec:discussion}

\subsection{Case Study}

\subsubsection{Case Study: 0x44c6c4}
Address 0x44c6c4~\cite{addr_0x44c6c4} represents the most active HFS swapper identified in our result. It executed a total of 4,469,893 swaps across 37 distinct token pairs, accumulating a notional trading volume of \$11.37B over 380~days of continuous activities. This corresponds to an average of 11,762 transactions per day and a daily traded volume of approximately \$29.9M. Such sustained intensity and temporal consistency throughout the year suggest a market-making or arbitrage engine continuously operating on decentralized exchanges rather than discretionary human trading.

\subsubsection{Case Study: 0x51c728}
We focus on swapper 0x51c728~\cite{addr_0x51C728}, which emerges as the second most active HFS swapper. Arkham Intelligence~\cite{arkm} tags this address as linked to `Wintermute: Market Maker', in which Wintermute~\cite {wintermute} is a leading institutional market-making and arbitrage firm. This attribution is consistent with prior academic studies. Both Yang et al.~\cite{yang2025decentralization} and Wu et al.~\cite{wu2025measuring} explicitly list this address among Wintermute-linked high-frequency searchers on Ethereum L1. From our extracted dataset, this address executed a total of 2,473,343 trades between 2023-09-14 14:13 and 2024-06-29 23:47, spanning 16 distinct token pairs and accumulating more than \$9.1B in notional volume.

\subsubsection{Case Study: 0xabdba4, 0x769d6a, and 0x2d512b}
A focused analysis of three HFS swappers, 0xabdba4~\cite{addr_0xabdba4}, 0x769d6a~\cite{addr_0x769D6a}, and 0x2d512b~\cite{addr_0x2d512B}, reveals strikingly consistent execution behavior that suggests coordinated related activity. Each address operates on a narrow set of token pairs, but large scale and with highly regular directional patterns. \textbf{a. }0x769d6a~\cite{addr_0x769D6a} is the most active among the group, maintaining uninterrupted swapping on the WETH/USDC from May~2023 to Apr.~2024, and later expanding its activity to the WETH/USDT0 market starting in Sep.~2023. Across both directions of these pairs, it executed 718,519 and 887,761 swaps, corresponding to notional volumes of \$2.47B and \$1.04B, respectively. \textbf{b. }0xabdba4~\cite{addr_0xabdba4} exhibits intensive activity on ARB/USDC, executing a total of 548,945 swaps with a combined notional volume of \$726.4M from May~2023 to Jan.~2024. \textbf{c. }0x2d512b~\cite{addr_0x2d512B} operates on the same ARB/USDC market but over a shorter time span, actively swapping 120,773 times with an aggregate notional volume of \$132.7M from Jun.~2023 to Aug.~2023. 

On-chain transfer data provide evidence of financial linkage among the three addresses. A total of 23 direct ERC-20 transfers were observed between them, primarily involving USDC and ARB. Most transfers involve sizable amounts ranging from 10,000 to 171,000 USDC, frequently clustered within Dec.~2023 and Jan.~2024, precisely overlapping the period of their intensive swapping. Notably, three transactions carry negligible values (0.001, 0.001, and 1 USDC), which are best interpreted as connectivity tests or wallet-link verifications, a behavior often observed in multi-account operational setups. 

\begin{figure}[t]
    \centering
    \includegraphics[width=\linewidth]{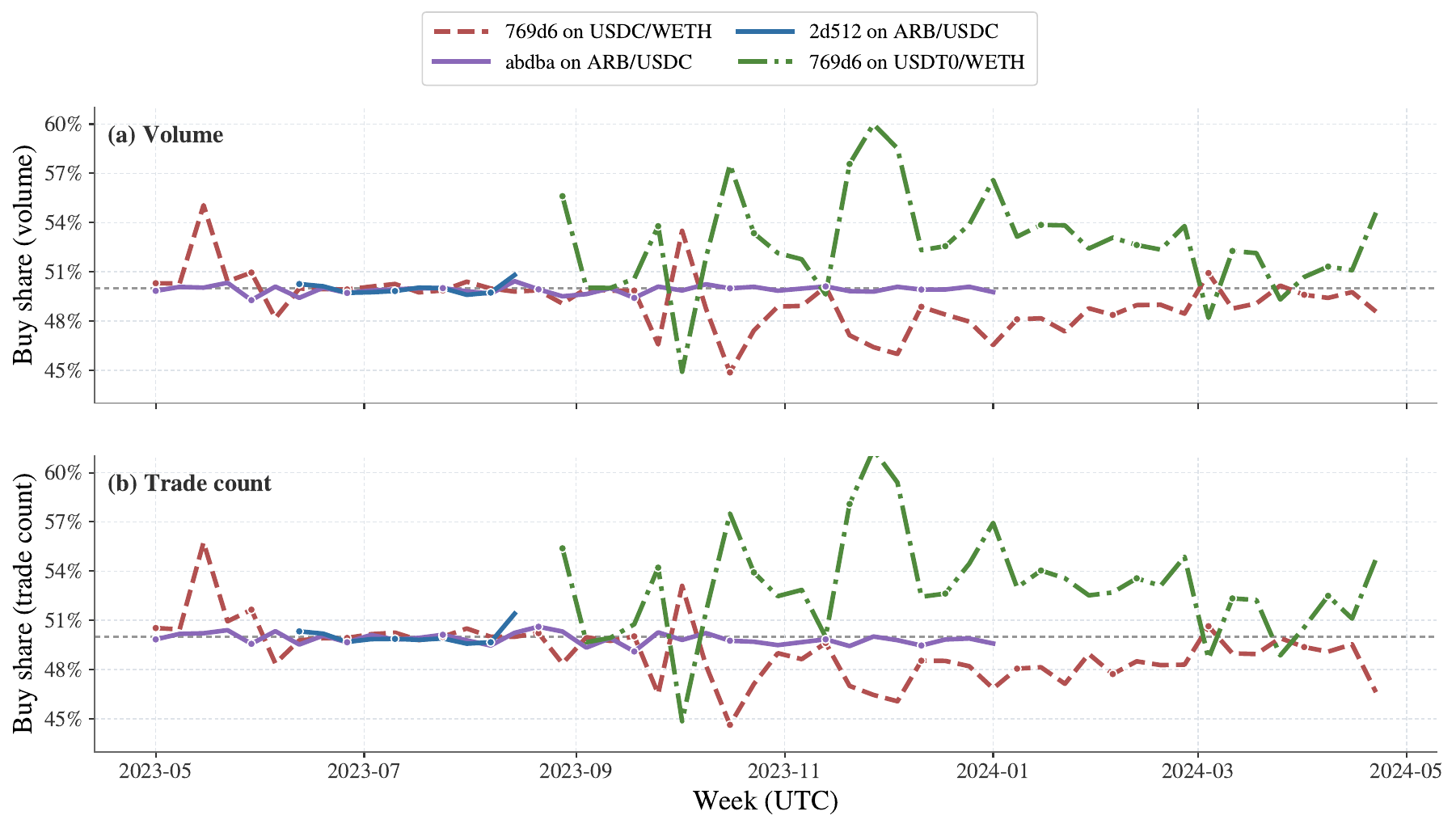}
    \caption{Daily buy-share by volume and by trade count for 0xabdba4, 0x769d6a, and 0x2d512b.}
    \label{fig:buy_share_symmetry}
\end{figure}

To quantify the degree of directional balance, we distinguish swaps that buy or sell the first symbol in each normalized trading pair (e.g., buying ARB in the ARB/USDC), and compare their respective frequencies and notional volumes across the three swappers. Figure~\ref{fig:buy_share_symmetry} illustrates their daily buy-share trajectories, with 50\% marking perfect symmetry between buying and selling. All three swappers exhibit persistent near-symmetric behavior, maintaining buy-shares between 45\% and 60\% over nearly their entire operational period. Aggregate statistics also reveal a consistent directional balance across all three HFS swappers. For instance, on USDC/WETH, 0x769d6a~\cite{addr_0x769D6a} executed 366,633 buy-side and 351,886 sell-side swaps with nearly equal notional volumes of \$1.26B and \$1.21B. A similar pattern appears on ARB/USDC for 0xabdba4~\cite{addr_0xabdba4} (275,164 vs.\ 273,781 trades, \$363.5M vs.\ \$362.8M) and for 0x2d512b~\cite{addr_0x2d512B} (60,385 vs.\ 60,388 trades, \$66.4M vs.\ \$66.3M). Such near-perfect symmetry in both trade count and notional value indicates that these entities employ programmatic, market-neutral strategies alternating between buy and sell operations with minimal directional bias. The alternation ratios remain consistently low, ranging from 0.12 for 0x769d6a~\cite{addr_0x769D6a} on USDC/WETH to 0.18 for the same address on USDT0/WETH, and 0.14–0.17 for 0x2d512b~\cite{addr_0x2d512B} and 0xabdba4~\cite{addr_0xabdba4} on ARB/USDC. This implies that only about one in six trades involves a direction switch. The median one-sided run length is short (typically 4 to 5 trades), but $P_{90}$ extends to 12–19 trades, and the longest continuous directional runs reach 100–200 consecutive swaps. Such persistence indicates that these three swappers do not alternate rapidly between buy and sell.

\subsection{Behavioral Signatures of HFS Swappers}
While on-chain data provides full visibility into transaction flows, it offers limited insight into the underlying intent or strategy of the actors involved. Without off-chain context, such as inventory information, one cannot fully reconstruct their decision logic. Nonetheless, by examining consistent temporal, directional, and pricing regularities, we can identify behavioral signatures suggestive of HFS activity.

\subsubsection{Stablecoin Arbitrage}
We isolate swaps between same-peg stablecoins (e.g., USDC, USDT). Among these, we identify 16,321 transactions where the execution rate was strictly above 1, meaning the swapper effectively received more than one unit of the bought stablecoin per unit of the sold one. Although each trade captures only a few basis points of edge, the aggregate effect is economically material, i.e., these micro-premia generated an estimated gross profit of about \$122.7K across \$168.1M volume. For example, on Jun 15, 2023, a HFS swapper executed a single swap~\cite{txn_0x64c56c} of \$750K USDC $\to$ USDC.e on Uniswap at a favorable rate of 1.000065, taking a profit of \$48.7. The existence of such a sizable subset of positive-edge trades demonstrates that HFS swappers might systematically harvest tiny inefficiencies in stablecoin pools. 


\subsubsection{CEX--DEX Exploration}
To investigate whether HFS swappers exploit price discrepancies between centralized exchange (CEX) and decentralized exchange (DEX), we construct a per-second \textit{CEX-DEX} execution edge that measures relative execution quality. For each token pair and swapper, all DEX transactions within the same second are aggregated to form a Volume-Weighted Average Price (VWAP)~\cite{mitchell2020volume}, denoted as \textit{$\mathrm{DEX}_t = \sum_i p_i q_i / \sum_i q_i$}, where \textit{$p_i$} and \textit{$q_i$} represent the price and traded quantity of the $i$-th swap in that second. The corresponding CEX reference price \textit{$\mathrm{CEX}_t$} is obtained from Binance spot market data via the public repository. The signed execution edge (in basis points) is then defined as \textit{$(\mathrm{CEX}_t-\mathrm{DEX}_t)/\mathrm{CEX}_t\times10^4$} when the swapper is a net buyer of the base token and \textit{$(\mathrm{DEX}_t-\mathrm{CEX}_t)/\mathrm{CEX}_t\times10^4$} when it is a net seller. Therefore, a positive value indicates a more favorable execution on DEXs, i.e., buying cheaper or selling higher than the simultaneous Binance quote. As a representative case study, we focus on the WETH--stablecoin pair on Arbitrum from Jun.~2023 to Jun.~2024. This one-year window covers periods of varying on-chain liquidity and market volatility, providing a robust basis for examining how DEX execution compares to centralized benchmarks under diverse regimes. To mitigate sampling noise from inactive or sporadic swappers, we exclude all \textit{(swapper, pool)} combinations with fewer than 1,000 observed seconds. After filtering, the dataset contains 124 high-quality combinations spanning three primary stablecoin markets (USDC/WETH, USDT0/WETH, DAI/WETH). Across these, the average mean execution edge is +5.4 bp, with an average of 69.6\% of seconds exhibiting positive DEX advantage, and a standard deviation of approximately 28 bp, indicating that, even at the second-by-second scale, DEX executions are, on average, marginally better than Binance spot, though substantial variation exists across swappers and pools.

\begin{figure}[t]
    \centering
    \includegraphics[width=0.95\linewidth]{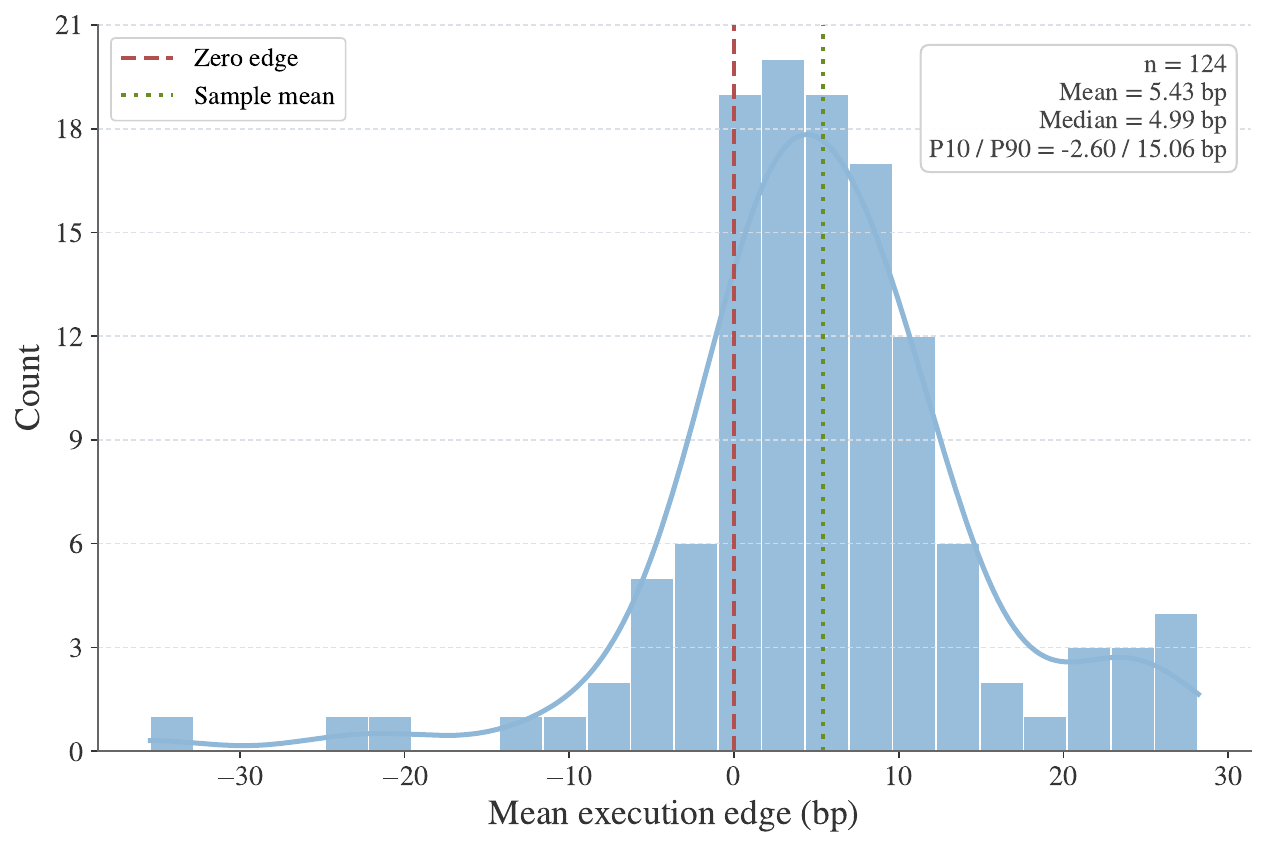}
    \caption{Distribution of mean execution edges.}
    \label{fig:edge_hist}
\end{figure}

\begin{figure}[t]
    \centering
    \includegraphics[width=0.95\linewidth]{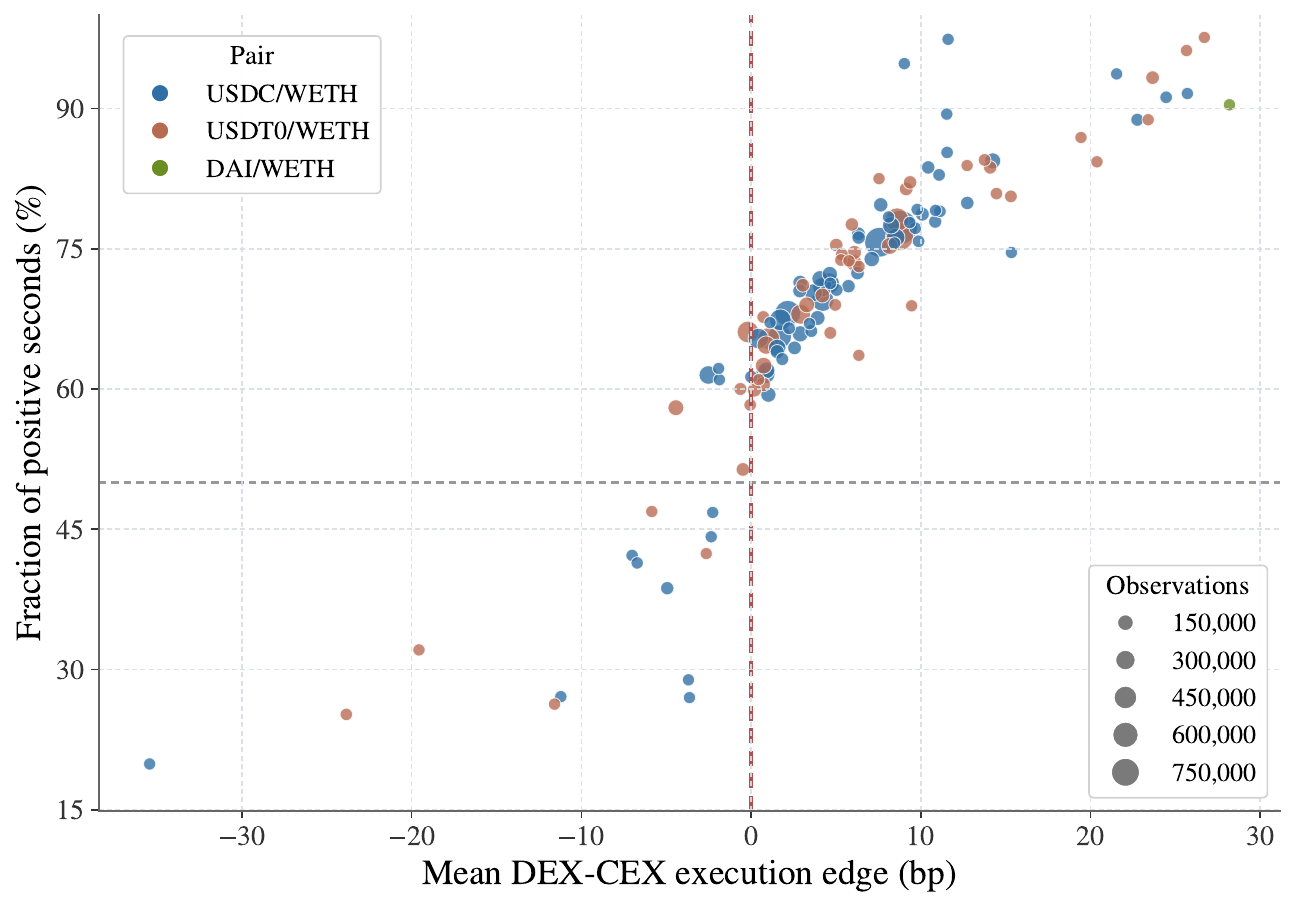}
    \caption{Cross-venue execution patterns.}
    \label{fig:edge_scatter}
\end{figure}

Figure~\ref{fig:edge_hist} displays the distribution of average execution edges across all qualified combinations. The distribution is right-skewed, peaking near +5~bp and extending up to +30~bp, indicating a mild but persistent DEX advantage in most cases. A minor left tail (down to –30~bp) corresponds to low-frequency or manual swappers experiencing systematically worse prices. Among them, the 0x98f989~\cite{addr_0x98f989} stands out as the most extreme case, with 2,011 observed seconds, an average edge of +26.9~bp, and 97.7\% of seconds showing a DEX-favorable execution. Complementing this, Figure~\ref{fig:edge_scatter} illustrates the cross-venue execution patterns. Each point represents a unique swapper–pair combination, with the x-axis showing mean execution edge, the y-axis showing the fraction of seconds with DEX-favorable pricing, and the marker size proportional to activity level. A strong positive correlation emerges, i.e., swappers with higher average edges are also exhibiting higher fractions of DEX-favorable seconds, suggesting consistent, rather than incidental, price advantages. Most swappers cluster around small positive edges (0–10~bp) and 60–80\% DEX-favorable seconds, while a small subset achieves edges above +20~bp and near-universal DEX advantage, likely corresponding to latency-optimized arbitrage bots.

Together, these findings reveal a heterogeneous but structurally consistent pattern of DEX price leadership, where a majority of active swappers achieve slight yet persistent execution advantages relative to centralized venues. However, unlike Ethereum L1, where proposer–builder separation and public mempools allow researchers to trace pre-trade intentions and builder-level collusion, the Arbitrum sequencer model provides no such transparency. We only observe executed swaps that have already been ordered and finalized on-chain, without visibility into the bidding, bundling, or transaction ordering process that might precede them. Consequently, any inferred cross-venue arbitrage mechanism remains a plausible, but not verifiable explanation.  Our analysis captures the statistical footprint of profitable latency behavior, but the strategic causality behind it remains opaque under L2’s single-sided observability.

\subsubsection{Round-Trip Trading}
To further characterize the behavioral patterns of HFS swappers, we analyze their \textit{round-trip trading} activities, using WETH–stablecoin pools as a representative case. We define a round-trip as a buy–sell (or sell–buy) sequence that occurs within a 24-hour window in the same liquidity pool. Specifically, for each swapper and trading pair, we first normalize all swaps into a consistent form, recording transaction time, direction (+1 for WETH purchases and –1 for WETH sales), execution price, and traded quantity. Each buy transaction is then matched to the earliest subsequent sell transaction by the same swapper in the same pool within 24 hours, forming a minimal \emph{round-trip} pair. The realized profit and return (in basis points) are computed as $\mathrm{PnL}_{i} = (p_{\mathrm{sell},i} - p_{\mathrm{buy},i}) \times q_{\mathrm{buy},i}$ and $r_{i} = \frac{p_{\mathrm{sell},i} - p_{\mathrm{buy},i}}{p_{\mathrm{buy},i}} \times 10^4$, respectively. This directional definition does not require matched quantities between legs, and each round-trip is measured relative to the notional exposure of the opening trade.  

We focus on transactions between Jun.~2023 and Jun.~2024, yielding all 161 swapper-pair samples covering 109 distinct swappers and two main stablecoin pairs (USDC/WETH and USDT0/WETH). The mean round-trip return is +5.7 bp (median +8.0 bp), and the weighted mean by pair count reaches +9.6 bp. Roughly 75\% of swapper-pair combinations exhibit positive average returns, suggesting mild but persistent profitability in short-term position reversals. The average volatility of round-trip returns is about 63 bp (weighted 42 bp), indicating significant heterogeneity across swappers. The two main stablecoin pools display comparable behavior, with average returns of +5.5 bp for USDC/WETH and +5.9 bp for USDT0/WETH. Notably, a small subset of addresses, including 0x6f15ee~\cite{addr_0x6f15ee}, 0x6117e2~\cite{addr_0x6117e2}, and 0x98f989~\cite{addr_0x98f989}, achieve average round-trip returns above +50 bp. While these results reveal short-term directional reversals and mild profitability, they do not imply on-chain self-containment of profit generation, e.g., off-chain hedging or inventory rebalancing on centralized venues may drive the observed on-chain cycles. Moreover, single-pool matching is a conservative lower bound, and round-trips can vanish at the pool level even when the strategy is active overall. For example, on ARB/USDC and ARB/USDT0, 0x51c728~\cite{addr_0x51C728} executed 303,082 swaps that were all buys of ARB with zero sells.

\subsection{Limitations}
Our analysis is subject to several limitations. First, on-chain traces capture executed outcomes but not the full trading context, such as off-chain hedging, inventory management, or internal risk transfers. Second, some measurements depend on operational thresholds (e.g., filtering rules and minimum observed seconds), so boundary cases may vary under alternative parameter settings. Third, the study window (Jan.~2023--Jun.~2024) and single-chain focus (Arbitrum) reflect a specific market regime, which limits immediate generalization to other periods or L2s. At the same time, we make best-effort design choices to improve reliability, including conservative filtering and multi-view empirical checks across temporal, directional, and cross-venue dimensions. Future work will extend the framework to additional L2s and market regimes, perform broader sensitivity analyses on key thresholds, and incorporate richer off-chain or cross-venue signals when available. Despite these constraints, the consistent temporal regularities and cross-venue relationships observed in our dataset support the robustness of the framework for characterizing HFS.


\section{Conclusion}
\label{sec:conclusion}
This paper presents the first systematic study of High-Frequency Swapping (HFS) on Arbitrum, based on 18 months of full-chain data. By isolating single-swap transactions and analyzing their temporal, directional, and cross-venue characteristics, we identify hundreds of persistent high-frequency actors responsible for tens of millions of swaps. The results reveal consistent machine-level regularity, balanced directional flows, and mild but measurable execution advantages relative to centralized venues. Together, these findings suggest that L2 rollups support a structured and automated form of short-horizon trading, even in the absence of public mempools or PBS, and establish a foundation for future research on decentralized market microstructure.

\bibliographystyle{splncs04}

\bibliography{refs}

@inproceedings{ferreira2024rolling,
  title={Rolling in the shadows: Analyzing the extraction of mev across layer-2 rollups},
  author={Ferreira Torres, Christof and Mamuti, Albin and Weintraub, Ben and Nita-Rotaru, Cristina and Shinde, Shweta},
  booktitle={Proceedings of the 2024 on ACM SIGSAC Conference on Computer and Communications Security},
  pages={2591--2605},
  year={2024}
}

@article{mitchell2020volume,
  title={Volume-weighted average price tracking: A theoretical and empirical study},
  author={Mitchell, Daniel and Bia{\l}kowski, Jedrzej and Tompaidis, Stathis},
  journal={IISE Transactions},
  volume={52},
  number={8},
  pages={864--889},
  year={2020},
  publisher={Taylor \& Francis}
}

@article{wu2025measuring,
  title={Measuring CEX-DEX Extracted Value and Searcher Profitability: The Darkest of the MEV Dark Forest},
  author={Wu, Fei and Sui, Danning and Thiery, Thomas and Pai, Mallesh},
  journal={arXiv preprint arXiv:2507.13023},
  year={2025}
}

@inproceedings{kalodner2018arbitrum,
  title={Arbitrum: Scalable, private smart contracts},
  author={Kalodner, Harry and Goldfeder, Steven and Chen, Xiaoqi and Weinberg, S Matthew and Felten, Edward W},
  booktitle={27th USENIX Security Symposium (USENIX Security 18)},
  pages={1353--1370},
  year={2018}
}

@article{wahrstatter2023time,
  title={Time to bribe: Measuring block construction market},
  author={Wahrst{\"a}tter, Anton and Zhou, Liyi and Qin, Kaihua and Svetinovic, Davor and Gervais, Arthur},
  journal={arXiv preprint arXiv:2305.16468},
  year={2023}
}

@article{malamud2017decentralized,
  title={Decentralized exchange},
  author={Malamud, Semyon and Rostek, Marzena},
  journal={American Economic Review},
  volume={107},
  number={11},
  pages={3320--3362},
  year={2017},
  publisher={American Economic Association 2014 Broadway, Suite 305, Nashville, TN 37203}
}

@article{hagele2024centralized,
  title={Centralized exchanges vs. decentralized exchanges in cryptocurrency markets: A systematic literature review},
  author={H{\"a}gele, Sascha},
  journal={Electronic Markets},
  volume={34},
  number={1},
  pages={33},
  year={2024},
  publisher={Springer}
}

@article{xu2023sok,
  title={Sok: Decentralized exchanges (dex) with automated market maker (amm) protocols},
  author={Xu, Jiahua and Paruch, Krzysztof and Cousaert, Simon and Feng, Yebo},
  journal={ACM Computing Surveys},
  volume={55},
  number={11},
  pages={1--50},
  year={2023},
  publisher={ACM New York, NY}
}

@inproceedings{yang2025decentralization,
  title={Decentralization of Ethereum's Builder Market},
  author={Yang, Sen and Nayak, Kartik and Zhang, Fan},
  booktitle={2025 IEEE Symposium on Security and Privacy (SP)},
  pages={1512--1530},
  year={2025},
  organization={IEEE}
}

@inproceedings{heimbach2023ethereum,
  title={Ethereum's Proposer-Builder Separation: Promises and Realities},
  author={Heimbach, Lioba and Kiffer, Lucianna and Ferreira Torres, Christof and Wattenhofer, Roger},
  booktitle={Proceedings of the 2023 ACM on Internet Measurement Conference},
  pages={406--420},
  year={2023}
}

@article{buterin2016ethereum,
  title={Ethereum: platform review},
  author={Buterin, Vitalik},
  journal={Opportunities and challenges for private and consortium blockchains},
  volume={45},
  pages={1--45},
  year={2016}
}

@inproceedings{wang2022cyclic,
  title={Cyclic arbitrage in decentralized exchanges},
  author={Wang, Ye and Chen, Yan and Wu, Haotian and Zhou, Liyi and Deng, Shuiguang and Wattenhofer, Roger},
  booktitle={Companion Proceedings of the Web Conference 2022},
  pages={12--19},
  year={2022}
}

@inproceedings{heimbach2022eliminating,
  title={Eliminating sandwich attacks with the help of game theory},
  author={Heimbach, Lioba and Wattenhofer, Roger},
  booktitle={Proceedings of the 2022 ACM on Asia Conference on Computer and Communications Security},
  pages={153--167},
  year={2022}
}

@article{oz2024wins,
  title={Who wins ethereum block building auctions and why?},
  author={{\"O}z, Burak and Sui, Danning and Thiery, Thomas and Matthes, Florian},
  journal={arXiv preprint arXiv:2407.13931},
  year={2024}
}

@inproceedings{zhou2021high,
  title={High-frequency trading on decentralized on-chain exchanges},
  author={Zhou, Liyi and Qin, Kaihua and Torres, Christof Ferreira and Le, Duc V and Gervais, Arthur},
  booktitle={2021 IEEE Symposium on Security and Privacy (SP)},
  pages={428--445},
  year={2021},
  organization={IEEE}
}

@inproceedings{li2023demystifying,
  title={Demystifying defi mev activities in flashbots bundle},
  author={Li, Zihao and Li, Jianfeng and He, Zheyuan and Luo, Xiapu and Wang, Ting and Ni, Xiaoze and Yang, Wenwu and Chen, Xi and Chen, Ting},
  booktitle={Proceedings of the 2023 ACM SIGSAC Conference on Computer and Communications Security},
  pages={165--179},
  year={2023}
}

@inproceedings{mclaughlin2023large,
  title={A large scale study of the ethereum arbitrage ecosystem},
  author={McLaughlin, Robert and Kruegel, Christopher and Vigna, Giovanni},
  booktitle={32nd USENIX Security Symposium (USENIX Security 23)},
  pages={3295--3312},
  year={2023}
}

@article{liu2025ethereum,
  title={Ethereum's Merge: Market liquidity, efficiency and volatility in the Proof of Stake Era},
  author={Liu, Bin and Prodromou, Tina and Suardi, Sandy and Xu, Caihong},
  journal={Economics Letters},
  volume={247},
  pages={112202},
  year={2025},
  publisher={Elsevier}
}

@inproceedings{moallemi2024analysis,
  title={An analysis of fixed-spread liquidation lending in defi},
  author={Moallemi, Ciamac and Patange, Utkarsh},
  booktitle={International Conference on Financial Cryptography and Data Security},
  pages={105--127},
  year={2024},
  organization={Springer}
}

@inproceedings{zhou2023sok,
  title={Sok: Decentralized finance (defi) attacks},
  author={Zhou, Liyi and Xiong, Xihan and Ernstberger, Jens and Chaliasos, Stefanos and Wang, Zhipeng and Wang, Ye and Qin, Kaihua and Wattenhofer, Roger and Song, Dawn and Gervais, Arthur},
  booktitle={2023 IEEE Symposium on Security and Privacy (SP)},
  pages={2444--2461},
  year={2023},
  organization={IEEE}
}

@inproceedings{yang2024sok,
  title={SoK: MEV countermeasures},
  author={Yang, Sen and Zhang, Fan and Huang, Ken and Chen, Xi and Yang, Youwei and Zhu, Feng},
  booktitle={Proceedings of the workshop on decentralized finance and security},
  pages={21--30},
  year={2024}
}

@inproceedings{qin2023mitigating,
  title={Mitigating decentralized finance liquidations with reversible call options},
  author={Qin, Kaihua and Ernstberger, Jens and Zhou, Liyi and Jovanovic, Philipp and Gervais, Arthur},
  booktitle={International Conference on Financial Cryptography and Data Security},
  pages={344--362},
  year={2023},
  organization={Springer}
}

@inproceedings{qin2022quantifying,
  title={Quantifying blockchain extractable value: How dark is the forest?},
  author={Qin, Kaihua and Zhou, Liyi and Gervais, Arthur},
  booktitle={2022 IEEE Symposium on Security and Privacy (SP)},
  pages={198--214},
  year={2022},
  organization={IEEE}
}

@inproceedings{qin2021empirical,
  title={An empirical study of defi liquidations: Incentives, risks, and instabilities},
  author={Qin, Kaihua and Zhou, Liyi and Gamito, Pablo and Jovanovic, Philipp and Gervais, Arthur},
  booktitle={Proceedings of the 21st ACM internet measurement conference},
  pages={336--350},
  year={2021}
}

@article{milionis2022automated,
  title={Automated market making and loss-versus-rebalancing},
  author={Milionis, Jason and Moallemi, Ciamac C and Roughgarden, Tim and Zhang, Anthony Lee},
  journal={arXiv preprint arXiv:2208.06046},
  year={2022}
}

@article{daian2019flash,
  title={Flash boys 2.0: Frontrunning, transaction reordering, and consensus instability in decentralized exchanges},
  author={Daian, Philip and Goldfeder, Steven and Kell, Tyler and Li, Yunqi and Zhao, Xueyuan and Bentov, Iddo and Breidenbach, Lorenz and Juels, Ari},
  journal={arXiv preprint arXiv:1904.05234},
  year={2019}
}

@inproceedings{chen2024exploring,
  title={Exploring the Security Issues of Real World Assets (RWA)},
  author={Chen, Shijian and Jiang, Muhui and Luo, Xiapu},
  booktitle={Proceedings of the Workshop on Decentralized Finance and Security},
  pages={31--40},
  year={2024}
}

@misc{l2beat_arb,
  author = {L2BEAT},
  title = {Arbitrum info \& stats in L2BEAT},
  note  = {\url{https://l2beat.com/scaling/projects/arbitrum}},
   year = 2025}

@misc{addr_0x2d512B,
  author = {Arbiscan},
  title = {Address: 0x2d512B},
  note  = {\url{https://arbiscan.io/address/0x2d512B5611a80810E3420619Ca9eE57d7d081F6b}},
   year = 2025}

@misc{addr_0x51C728,
  author = {Arbiscan},
  title = {Address: 0x51C728},
  note  = {\url{https://arbiscan.io/address/0x51C72848c68a965f66FA7a88855F9f7784502a7F}},
   year = 2025}

@misc{addr_0xabcd71,
  author = {Arbiscan},
  title = {Address: 0xabcd71},
  note  = {\url{https://arbiscan.io/address/0xabcd71f0330c011260d79f19c147a95dc75f7975}},
   year = 2025}

@misc{addr_0x769D6a,
  author = {Arbiscan},
  title = {Address: 0x769D6a},
  note  = {\url{https://arbiscan.io/address/0x769D6a494aEF318B175b76eC19dcBef3AC6f4A05}},
   year = 2025}

@misc{addr_0xE27AE1,
  author = {Arbiscan},
  title = {Address: 0xE27AE1},
  note  = {\url{https://arbiscan.io/address/0xE27AE153238A965df073c1eeECa9860829efaa47}},
   year = 2025}

@misc{addr_0x98f989,
  author = {Arbiscan},
  title = {Address: 0x98f989},
  note  = {\url{https://arbiscan.io/address/0x98f9894a4768256b7092d8a975a1ac13dd7a9f01}},
   year = 2025}

@misc{addr_0x44c6c4,
  author = {Arbiscan},
  title = {Address: 0x44c6c4},
  note  = {\url{https://arbiscan.io/address/0x44c6c4c7da174f47ac81e818d55c02600390e79d}},
   year = 2025}

@misc{addr_0xabdba4,
  author = {Arbiscan},
  title = {Address: 0xabdba4},
  note  = {\url{https://arbiscan.io/address/0xabdba477f4111ce4a1e1ad0f6b0a4593d45c29c4}},
   year = 2025}

@misc{addr_0x6f15ee,
  author = {Arbiscan},
  title = {Address: 0x6f15ee},
  note  = {\url{https://arbiscan.io/address/0x6f15ee9258acdebf356db7ab607bb255a00c6fdf}},
   year = 2025}

@misc{addr_0x6117e2,
  author = {Arbiscan},
  title = {Address: 0x6117e2},
  note  = {\url{https://arbiscan.io/address/0x6117e2b93db124c1c6248168e56ac3b22a88779a}},
   year = 2025}

@misc{pool_0xb0f6ca,
  author = {GeckoTerminal},
  title = {ARB/USDC on Uniswap V3 (Arbitrum)},
  note  = {\url{https://www.geckoterminal.com/arbitrum/pools/0xb0f6ca40411360c03d41c5ffc5f179b8403cdcf8}},
   year = 2025}

@misc{token_usdce,
  author = {Arbiscan},
  title = {Token Bridged USDC (USDC.e)},
  note  = {\url{https://arbiscan.io/token/0xff970a61a04b1ca14834a43f5de4533ebddb5cc8}},
   year = 2025}

@misc{token_usdc,
  author = {Arbiscan},
  title = {Token USD Coin (USDC)},
  note  = {\url{https://arbiscan.io/token/0xaf88d065e77c8cc2239327c5edb3a432268e5831}},
   year = 2025}

@misc{txn_0x91601f,
  author = {Arbiscan},
  title = {Transaction: 0x91601f...21b74a},
  note  = {\url{https://arbiscan.io/tx/0x91601f0c50092f355914a2960037961e85badadb4887d84061f89ac04221b74a}},
   year = 2025}

@misc{txn_0x9517a1,
  author = {Arbiscan},
  title = {Transaction: 0x9517a1...6d6d03},
  note  = {\url{https://arbiscan.io/tx/0x9517a10708419851ba3d0894925f3893fea79ef7f6e7b5963e020fdc546d6d03}},
   year = 2025}

@misc{txn_0x1e9728,
  author = {Arbiscan},
  title = {Transaction: 0x1e9728...bb698b},
  note  = {\url{https://arbiscan.io/tx/0x1e972887cb3f185c69ba4d78e32df4aa4f35c83bb8507ddde11aa06a67bb698b}},
   year = 2025}

@misc{txn_0x7e56f3,
  author = {Arbiscan},
  title = {Transaction: 0x7e56f3...a57903},
  note  = {\url{https://arbiscan.io/tx/0x7e56f32e7adf32a90ea5f1f5e80698d0d26a6b909822b394474bf9206aa57903}},
   year = 2025}

@misc{txn_0xb1f880,
  author = {Arbiscan},
  title = {Transaction: 0xb1f880...a4c379},
  note  = {\url{https://arbiscan.io/tx/0xb1f8800fa3da1a1fd6abbf0325f4df083088c3c177af18da34d44ce433a4c379}},
   year = 2025}

@misc{txn_0x64c56c,
  author = {Arbiscan},
  title = {Transaction: 0x64c56c...24719a},
  note  = {\url{https://arbiscan.io/tx/0x64c56ccc636a49c2e25f5d9f1f8ea4fedc7acf3bc43d696508056bddcc24719a}},
   year = 2025}

@misc{arkm,
  author = {Arkham},
  title = {Arkham Intelligence},
  note  = {\url{https://intel.arkm.com/}},
   year = 2025}

@misc{wintermute,
  author = {Wintermute},
  title = {Harnessing chaos in digital asset markets. Wintermute makes digital asset markets liquid and efficient},
  note  = {\url{https://www.wintermute.com/}},
   year = 2025}

@misc{flowtraders,
  author = {FLOW TRADERS},
  title = {A global trading firm driving transparency and efficiency in financial markets},
  note  = {\url{https://www.flowtraders.com/}},
   year = 2025}

@misc{selini,
  author = {Selini Capital},
  title = {Selini is a global trading firm focused on systematic trading and venture investing
- all within the digital assets space.},
  note  = {\url{https://www.selinicapital.com/}},
   year = 2025}

@misc{manifold,
  author = {Manifold Trading},
  title = {Manifold is a systematic, 
quantitative investment firm},
  note  = {\url{http://manifoldtrading.com/}},
   year = 2025}

@misc{mevboost,
  author = {MEV-Boost},
  title = {MEV-Boost Dashboard},
  note  = {\url{https://mevboost.pics/}},
   year = 2025}

@misc{mevboost1,
  author = {MEV-Boost},
  title = {mev-boost is open source middleware run by validators to access a competitive block-building market.},
  note  = {\url{https://github.com/flashbots/mev-boost}},
   year = 2025}

@article{fano1947ionization,
  title={Ionization yield of radiations. II. The fluctuations of the number of ions},
  author={Fano, Ugo},
  journal={Physical Review},
  volume={72},
  number={1},
  pages={26},
  year={1947},
  publisher={APS}
}

@inproceedings{milionis2024automated,
  title={Automated market making and arbitrage profits in the presence of fees},
  author={Milionis, Jason and Moallemi, Ciamac C and Roughgarden, Tim},
  booktitle={International Conference on Financial Cryptography and Data Security},
  pages={159--171},
  year={2024},
  organization={Springer}
}

\end{document}